\documentclass[12pt,preprint]{aastex}
\usepackage{color}

\newcommand{\msun}{{\rm\,M_\odot}}
\newcommand{\Msun}{{\msun}}

\newcommand{\ct}{\citealt}

\begin{document}


\title{Magnetic Flux Expulsion in Star Formation} 

%
%
\author{Bo Zhao\altaffilmark{1}, Zhi-Yun Li\altaffilmark{1}, Fumitaka
  Nakamura\altaffilmark{2}, Ruben Krasnopolsky\altaffilmark{3}, Hsien Shang\altaffilmark{3}}
\altaffiltext{1}{University of Virginia, Astronomy Department, Charlottesville, USA}
\altaffiltext{2}{NAOJ, Japan} 
\altaffiltext{3}{Academia Sinica, Institute of Astronomy and Astrophysics, Taipei, Taiwan}
\shortauthors{{\sc Zhao et al.}}
\shorttitle{{\sc Magnetic Flux Expulsion in Star Formation}}

\begin{abstract}

Stars form in dense cores of magnetized molecular clouds. If the
magnetic flux threading the cores is dragged into the stars, the
stellar field would be orders of magnitude stronger than observed. 
This well-known 
``magnetic flux problem'' demands that most of the core magnetic flux 
be decoupled from the matter that enters the star.
We carry out the first exploration of what happens to the decoupled
magnetic flux in 3D, using an MHD version of the ENZO adaptive mesh 
refinement code. The field-matter decoupling is achieved
through a sink particle treatment, which is needed to follow the
protostellar accretion phase of star formation. We find that the
accumulation of the decoupled flux near the accreting protostar 
leads to a magnetic pressure buildup. The high pressure is released 
anisotropically, along the path of least resistance. It drives a
low-density expanding region in which the decoupled magnetic flux 
is expelled. This decoupling-enabled magnetic structure has
never been seen before in 3D MHD simulations of star formation. It
generates a strong asymmetry in the protostellar accretion flow,
potentially giving a kick to the star. In the presence of an initial
core rotation, the structure presents an obstacle to the formation 
of a rotationally supported disk, in addition to magnetic braking, 
by acting as a rigid magnetic wall that prevents the rotating 
gas from completing a full orbit around the central object. We 
conclude that the decoupled magnetic flux from the stellar matter 
can strongly affect the protostellar collapse dynamics.

\end{abstract}

\section{Introduction}  
\label{intro}

%
%

A longstanding problem in star formation is the so-called ``magnetic 
flux problem'' (e.g., \ct{Nakano1984}, see his \S~4). If the magnetic 
flux observed to thread a typical star-forming dense core
(\ct{TrolandCrutcher2008}) were to be dragged into a young stellar
object, the stellar field strength would be tens of millions of Gauss,
more than three orders of magnitude higher than the observed values
(which are typically in the kilo-Gauss range, e.g.,
\ct{Johns-Krull2009}). The vast majority of the original magnetic flux
of the dense core must be decoupled from the matter that enters the
star. When and how the decoupling occurs is a fundamental problem of
star formation that has yet to be completely resolved. 


The magnetic flux problem lies at the heart of another fundamental 
problem in star formation: disk formation. If the magnetic flux of
a dense core is dragged by the collapsing material into a forming 
star, as would be in the ideal MHD limit, it would form a central 
split magnetic monopole that prevents the formation of a rotationally 
supported disk through catastrophic magnetic braking (\ct{Allen+2003}; 
\ct{Galli+2006}; \ct{MellonLi2008}; \ct{HennebelleFromang2008}). The 
magnetic flux problem must be resolved in order for rotationally 
supported disks to form. 


The most widely discussed resolution of the magnetic flux problem is
through non-ideal MHD effects, including ambipolar diffusion, Ohmic
dissipation and potentially Hall effect (e.g., \ct{Nakano1984};
\ct{LiMcKee1996}; \ct{CiolekKonigl1998}; \ct{TassisMouschovias2007}; \ct{KunzMouschovias2010}; \ct{Krasnopolsky+2011}). For example, \citet{LiMcKee1996} showed that ambipolar diffusion can enable the protostellar envelope to collapse into the central stellar object without dragging along the magnetic flux. The left-behind magnetic flux builds up in a small circumstellar region, confined by the ram pressure of the protostellar collapse. It tends to dominate the gas dynamics close to the protostar, particularly in the region of disk formation. Indeed, the magnetic field accumulated at small radii can be strong enough to suppress disk formation completely through efficient magnetic braking (\ct{KrasnopolskyKonigl2002}; \ct{MellonLi2009}; \ct{Li+2011}). In this case, the resolution of the magnetic flux problem for the central star does not lead to a resolution of the magnetic braking problem for disk formation.

There is, however, a significant limitation in the (non-ideal MHD)
studies of the magnetic flux problem and its consequences to date: the
assumption of axisymmetry. Although the axisymmetry greatly reduces
the computational demand of the calculations, it limits how fast the
magnetic flux released from the central object can expand to large
distances. In particular, it suppresses a likely mode for the flux
expulsion: dynamic expansion along the direction(s) of least
resistance. In this paper, we carry out the first detailed 3D study of
what happens to the released flux using the Enzo MHD code. We find
that a magnetically dominated region is inflated by the released
flux. The region expands asymmetrically away from the central object,
changing the dynamics of the protostellar accretion and disk formation. 

The rest of the paper is organized as follows. In \S~\ref{setup}, we
describe the problem setup, including the equations to be solved,
numerical method, and initial and boundary conditions. The numerical
results are presented and interpreted in \S~\ref{result}. The last
section, \S~\ref{discussion}, includes a discussion of the main
results and a short summary. 

\section{Problem Setup}
\label{setup}

\subsection{Basic Equations and Numerical Method}
\label{method}

We study the formation of stars from the collapse of magnetized dense 
cores of molecular clouds using an MHD code that includes a sink
particle treatment. The usual MHD equations are the magnetic
induction equation, 
\begin{equation}
{\partial {\bf B}\over \partial t} = \nabla \times ({\bf v}\times {\bf B}),
\end{equation}
and the equations for mass continuity, momentum and self-gravity,

\begin{equation}
{\partial \rho \over \partial t} + \nabla \cdot (\rho {\bf v}) = 0,
\end{equation}

\begin{equation}
\rho {\partial {\bf v} \over \partial t} + \rho({\bf v} \cdot \nabla) {\bf v} = -\nabla P - {1 \over 4\pi}{\bf B} \times (\nabla \times {\bf B}) - \rho \nabla \phi,
\end{equation}

\begin{equation}
{\nabla}^2 \phi = 4 \pi G \rho,
\end{equation}
where $\phi$ is the gravitational potential and other symbols have their usual meanings. 

The above equations
are solved in three dimensions using an MHD
version (\ct{WangAbel2009}) of the ENZO adaptive mesh refinement code
(\ct{BryanNorman1997}; \ct{O'Shea+2004}). It 
incorporates a sink particle treatment (\ct{Wang+2010}). The magnetic
field 
is evolved with a conservative MHD solver that includes the hyperbolic
divergence cleaning of \citet{Dedner+2002}. The MHD version of the 
code is publicly available from the ENZO website at {\it http://code.google.com/p/enzo/}. It has been used to 
follow successfully the formation and evolution of magnetized galaxies 
(\ct{WangAbel2009}) and star clusters (\ct{Wang+2010}). 

Our goal is to follow both the prestellar core evolution as well as
the protostellar mass accretion phase of star formation (after a 
central stellar object has formed). For the latter, it is crucial to use a sink
particle to approximate the stellar object, because including the 
object in the computation without any special treatment would reduce the time step to such a small value that the 
simulation would grind to a halt (e.g., \ct{Krumholz+2004}). 
In our simulations, we resolve the so-called ``Jeans length,'' defined
as  $L_J=c_s\left({\pi\over G\rho}\right)^{1/2}$ (where $c_s$ is the isothermal
sound speed), everywhere by at least 8 cells, so that the Truelove's criterion
(\ct{Truelove+1997}) is satisfied. When the density in a cell at the
highest refinement level exceeds the threshold density $\rho_{th} =
{5\pi c_s^2 \over 3G (8\Delta x)^2}$ (where $\Delta x$ is the size of the
smallest cell), a sink particle is created at the cell center. The 
particle is evolved using an algorithm that is described in detail 
in \citet{Wang+2010}. 

Briefly, the mass accretion rate onto a sink particle is done in two
steps. First, the particle accretes from its host cell using a formula
inspired by that of Bondi-Hoyle accretion (\ct{Ruffert1994}).
%
%
The momentum of the accreted material is added to that of the sink
particle. The second step involves the merging of small sink 
particles. This step is controlled
by two parameters: the merging mass $M_{merg}$ and merging distance
$l_{merg}$. They are chosen to eliminate artificial particles and 
to maximize computation efficiency. The simulations presented in this
paper use $M_{merg} = 0.01 \msun$ and $l_{merg} = 10^{15} cm \approx
8{\Delta}x \approx 70 AU$. Our main results are insensitive to these 
parameters as long as they are reasonably small.   

When a mass is extracted from a cell, due to either sink particle
formation or accretion onto an existing sink, the magnetic field
in that cell is not altered. That is, the field strength remains the
same. This is a crude way to represent the decoupling of the magnetic
field from matter at high densities that is expected physically and
demanded by the relatively weak magnetic fields observed on young
stars (see discussion in 
\S~\ref{intro}). Determining what happens to the decoupled magnetic 
flux in 3D is the main goal of our investigation. 

\subsection{Initial and Boundary Conditions}
\label{initialBC}


We model star formation in a magnetized dense core embedded in a more
diffuse ambient medium. We choose a spherical core of radius $R=5\times 10^{16}$~cm and an initially uniform density
$\rho_0=5\times 10^{-18}$~g~cm$^{-3}$. These parameters yield a core mass $M=1.32$~$\Msun$ and  
a free-fall time $t_{ff}=9.4\times 10^{11}$~sec = 29.8~kyrs. We embed
the core in an ambient medium that is 100 times less dense than the
core and that fills the entire computational box, which is much larger than the core, with $L=5\times 10^{17}$~cm on each
side. The large box size is chosen to minimize the effects of the
periodic boundary conditions on the core dynamics; the conditions are
adopted to facilitate the computation of self-gravity (through fast
Fourier transform on the base grid). 
%


%
%

As usual, we adopt a barotropic equation of state (EOS) that mimics
the isothermal EOS at low densities
and the adiabatic EOS at high densities: 
\begin{equation}
P = \rho c_s^2 \left[1 + \left({\rho\over\rho_{crit}}\right)^{2/3}\right],
\end{equation}
with a critical density 
$\rho_{crit}=10^{-13}$~g~cm$^{-3}$ for the transition between the two
regimes. An isothermal sound speed
$c_s=0.2$~km/s is chosen, corresponding to a temperature $T\sim 10$~K. The
sound speed yields a ratio of thermal to gravitational
energy $\alpha=2.5Rc_s^2/(GM)=0.29$ for the core. For simplicity, we
impose an initially uniform magnetic field everywhere along the
$z-$axis, with a strength $B_0=2.7\times 10^{-4}$~G. It corresponds to a
dimensionless mass-to-flux ratio $\lambda_{core} = 2$ for the core as
a whole, in units of the critical value 
%
%
$(2{\pi}G^{1/2})^{-1}$; such a value is consistent with those predicted in
dense cores formed in strongly magnetized clouds through ambipolar
diffusion (e.g., \ct{LizanoShu1989}; \ct{BasuMouschovias1994};
 \ct{NakamuraLi2005}), and with the median value inferred by
\citet{TrolandCrutcher2008} for a sample of dark cloud cores (after
correcting statistically for projection effects). We have also carried
simulations for both weaker and stronger field cases, with
$\lambda_{core}=4$ and $1$ respectively, and found qualitatively similar
results. We should note that the local mass-to-flux ratio on the
central magnetic flux tube is somewhat larger than the value for the
core as a whole, by a factor of 1.5. 

Besides the magnetic field, we impose, in some cases, a rigid body
rotation on the core. We adopt an angular velocity of $\Omega=4.0
\times 10^{-13}$~s$^{-1}$, corresponding to a ratio of rotational and
gravitational energy $\beta= 0.036$. This value is within the range
inferred by \citet{Goodman+1993} from NH$_3$ observations of dense cores. 


We choose a relatively coarse base grid of $64^3$, although the grid is
automatically refined, at the beginning of the simulation, by one 
level in the central part of the computational domain that includes 
the dense core. We set the maximum refinement level to 6 for our
reference runs, which yields 
a smallest cell size of $\Delta x \sim 8$~AU. 

\section{Results}
\label{result}

We carry out two sets of simulations: collapse with or without
rotation. The former is to illustrate how the magnetic flux decoupled
from the stellar material escapes to large distances and the latter
the effects of the escaping flux on disk formation. 

\subsection{Non-Rotating Collapse}
\label{non-rotating}

\subsubsection{Sink Particle Evolution}

At the heart of our star formation calculation lies the sink particle
treatment. As mentioned in \S~\ref{method}, the treatment is needed to
allow the simulation to go beyond the initial prestellar core
evolution phase of star formation, into the protostellar mass 
accretion phase. Just as 
importantly, it provides a simple way to decouple the magnetic flux
from the material that enters the protostellar object, as demanded by
observations (see discussion in \S~\ref{intro}). 

In the left panel of Fig.~1, we plot the mass of the star, as
represented by the sink particle, as a function of time. The object 
first appears around $t \approx 36$~kyrs (or $\sim 1.2\ t_{ff}$). It grows
quickly, with a relatively large initial mass accretion rate 
of $10^{-4} \msun$~yr$^{-1}$. The high initial accretion rate is
expected for the uniform density distribution that we
adopted for the dense core. It could plausibly be identified with the
Class 0 phase of low-mass star formation (\ct{Andre+1993}). The accretion
rate decreases below $10^{-5}\msun$~yr$^{-1}$ after $t = 48$~kyrs (or
$\sim 1.6~t_{ff}$), when $\sim 0.75 \msun$ (or $57\%$ of the initial 
core mass) has landed on the central star. 

\begin{figure}
\epsscale{1.2}
\plottwo{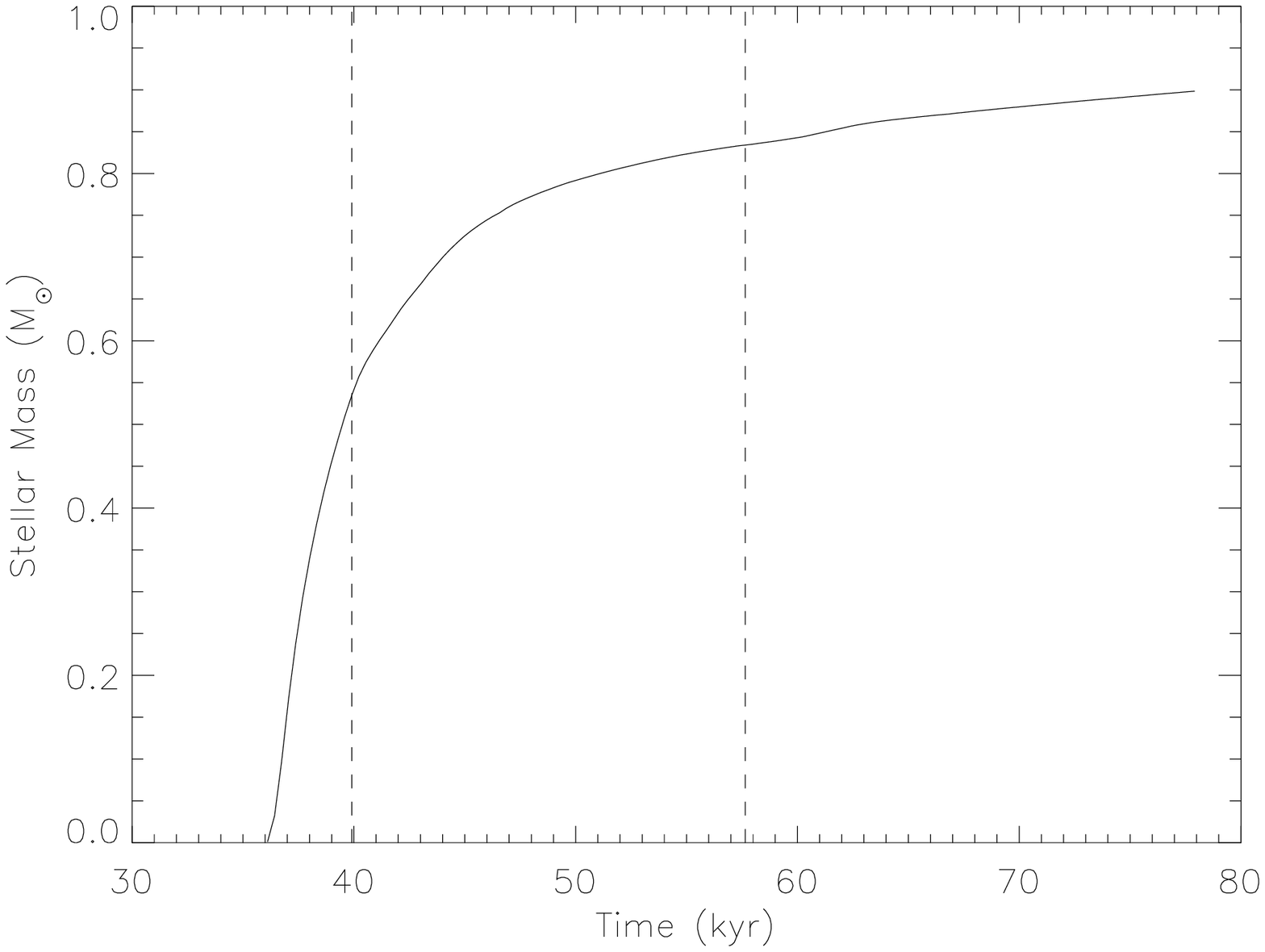}{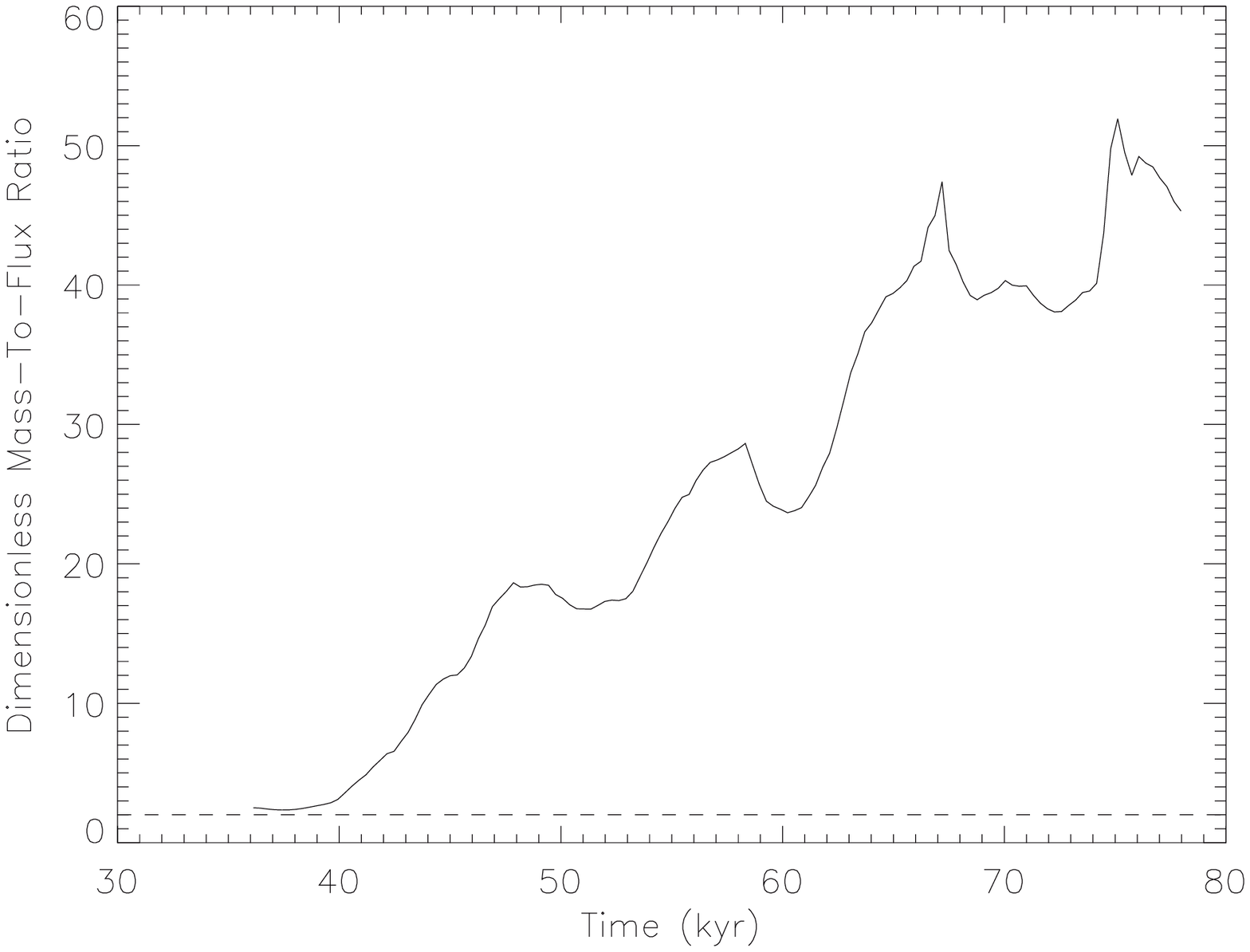}
\caption{Left panel: Mass of the protostar (as represented by a sink
  particle) as a function of time. The dotted vertical lines denote the
  range of time plotted in Fig.~\ref{mtf2}. Right panel: Dimensionless
  mass-to-flux ratio within 200~AU of the star, indicating that most
  of the magnetic flux associated with the stellar mass is left
  outside of the small region (see text). The initial core
  mass-to-flux ratio $\lambda_{core}=2$ is plotted (dashed line) for comparison.}
\label{SF}
\end{figure} 

To make sure that the sink particle treatment has indeed decoupled the
magnetic flux from the material that enters the stellar object,
we plot in the right panel of Fig.~\ref{SF} the dimensionless ratio 
$\lambda$ of all mass (including sink) and all magnetic flux 
within a small radius (200~AU) around the sink particle. The flux is
computed on the x-y plane that passes through the star (i.e., the constant
$z=z_{\star}$ plane where $z_{\star}$ is the stellar position in 
the $z-$direction); this plane will be referred to as the equatorial plane
of the star or equatorial plane for short hereafter. There is a
general trend for $\lambda$ to increase
with time, reaching values as high as 50, which is much higher than
the dimensionless mass-to-flux ratio for the core as a whole
($\lambda_{core}=2$). Clearly, the magnetic flux near the protostar
did not increase as fast as the stellar mass, and the vast majority of
the flux originally associated with the stellar mass must reside 
outside the small region. It is natural to ask: where did this flux
go?

\subsubsection{Decoupling-Enabled Magnetic Structure}

\begin{figure}
\epsscale{1.0}
\plottwo{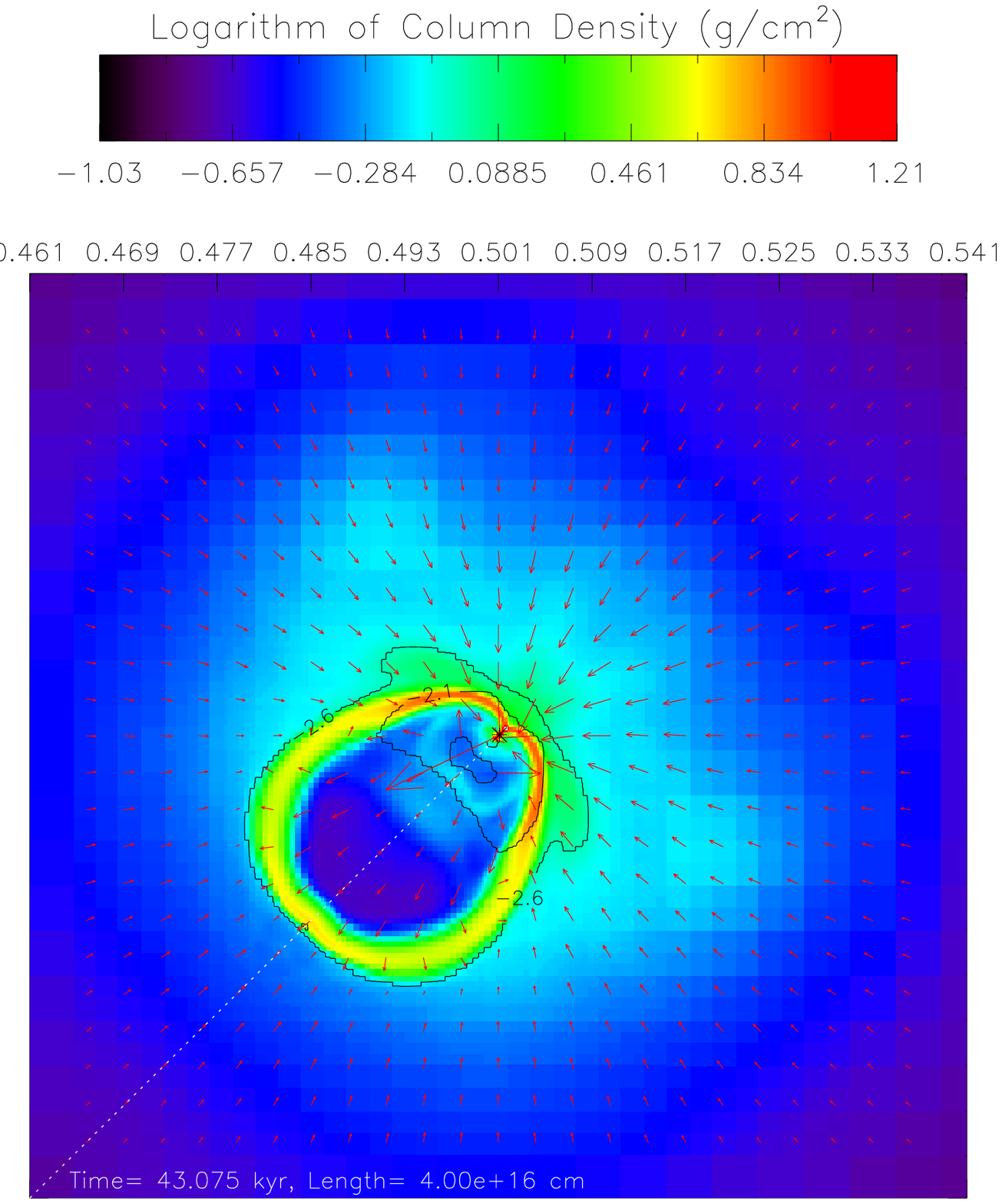}{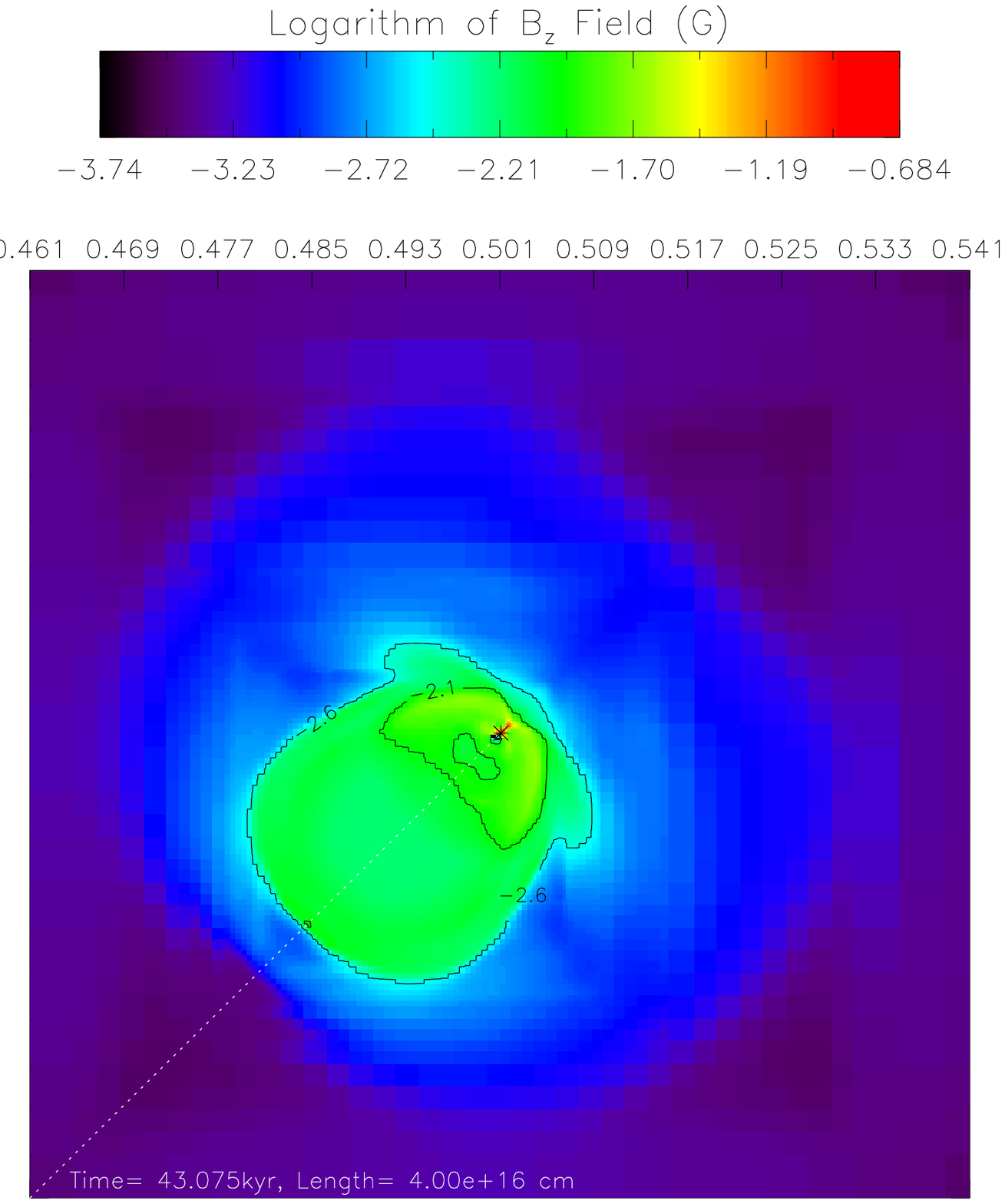}
\caption{Left panel: Column density (along $z-$direction) and
  velocity field (on the equatorial plane) of the inner region of the
  collapsing core at a
  representative time $t=43$~kyrs (or $\sim 1.45\ t_{ff}$), showing an
  expanding, evacuated region to the lower-left of the star (marked by
  an asterisk on the map). Superposed on the map are contours of
  constant $B_z$ (the z-component of the magnetic field) on the
  equatorial plane. Right panel: Color map of $B_z$ on the equatorial
  plane, showing that the evacuated region coincides with a region of
  intense magnetic field. The tick marks above each panel are in units of 
  $5 \times 10^{17}$ cm (size of the simulation box).
}
\label{magbub}
\end{figure}

It turns out that the decoupled flux is trapped in a strongly
magnetized, low-density structure that expands with time. The left
panel of Fig.~\ref{magbub} shows
the structure in a map of column density (along $z-$direction) at a
representative time $t=43$~kyrs (or $\sim 1.45\ t_{ff}$). At this time,
the evacuated region has a size of $\sim 1.2 \times 10^{16}$~cm (or
about 800~AU). It grows in time, as indicated by the velocity vectors
inside the region. The right panel of Fig.~\ref{magbub} plots the $z-$component
of the magnetic field, $B_z$, on the equatorial plane of the
star. It shows that the hollow region coincides with a region of
intense magnetic field, leaving little doubt that the structure has a
magnetic origin. 

\begin{figure}
\epsscale{0.8}
\plotone{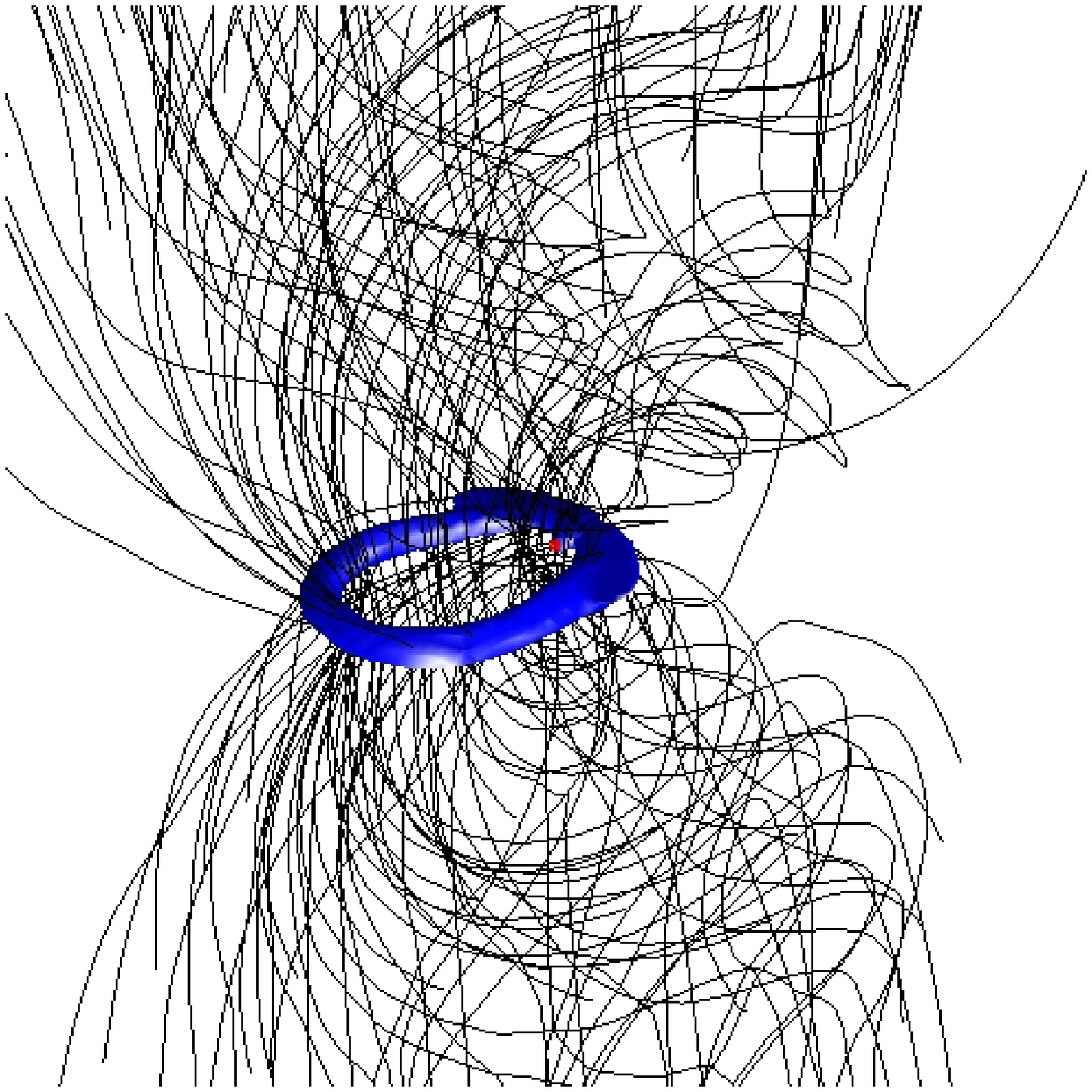}
\caption{3D view of the inner part of the collapsing core, showing the
  decoupling-enabled magnetic structure (DEMS), where a bundle of
  twisted magnetic field lines is surrounded at the ``waist'' by a
  dense ring. The star is shown as a red dot located near the inner 
edge of the ring.}
\label{twist}
\end{figure}

One may be tempted to call the low-density structure a ``magnetic
bubble.'' However,
as shown in Fig.~\ref{twist}, ``bubble'' does not provide an adequate
description of the structure in 3D; the dense feature surrounding 
the evacuated region in the column density map (Fig.~\ref{magbub}) turns out 
to be a ring rather than a shell. The reason is that, in the absence
of the decoupled magnetic flux, the collapsing core material would
settle preferentially along field lines into a dense, flatten,
structure---a ``pseudodisk'' (\ct{GalliShu1993}). The left-behind magnetic
flux evacuates part of the pseudodisk, creating the structure showing
in Fig.~3, where a bundle of magnetic field lines is pinched near the
equator by a dense ring. We will refer to the structure as the
``decoupling-enabled magnetic structure'' (DEMS for short hereafter). 

The strongest evidence for the DEMS being really driven by the magnetic
flux decoupled from the material accreted onto the stellar object 
comes from Fig.~4, where we plot the ratio of the stellar mass $M_{\star}$
to the magnetic flux in the structure $\Phi_d$. The quantity $\Phi_d$
is computed on the equatorial plane of the star. It includes
all cells inside a boundary beyond which $B_z$ falls off steeply (see Fig.~\ref{magbub}). Not 
surprisingly, the cells are mostly located inside the dense
ring. We are able to determine the ratio over only a limited range in time, 
because the DEMS becomes apparent only after $t \sim  40$~kyrs, and it
starts to merge into the background (making an accurate identification of
the DEMS difficult) after $t \sim 60$~kyrs. As the stellar mass nearly
doubles during
the early part of this period (see Fig.~\ref{SF}), the mass-to-flux ratio
remains relatively constant, indicating that $\Phi_d$ increases
together with $M_{\star}$. This behavior is consistent with $\Phi_d$
being released by $M_{\star}$. Furthermore, the dimensionless
mass-to-flux ratio $\lambda_{\star,d} = 2\pi G^{1/2} M_{\star}/
\Phi_d$ is close to the global value for the initial core, as one
would expect if the initial core matter releases its flux on the way
to the center. The small deviation from the global value $\lambda_{core}=2.0$
can come from the fact that local $\lambda$ in the initial core is not
exactly 2, and that the DEMS is identified by eye and is not very
precise. Nevertheless, the different pieces of evidence presented in
this subsection leave little doubt that the DEMS is formed by the
magnetic flux decoupled from the accreted stellar mass. 

\begin{figure}
\epsscale{1.0}
\plotone{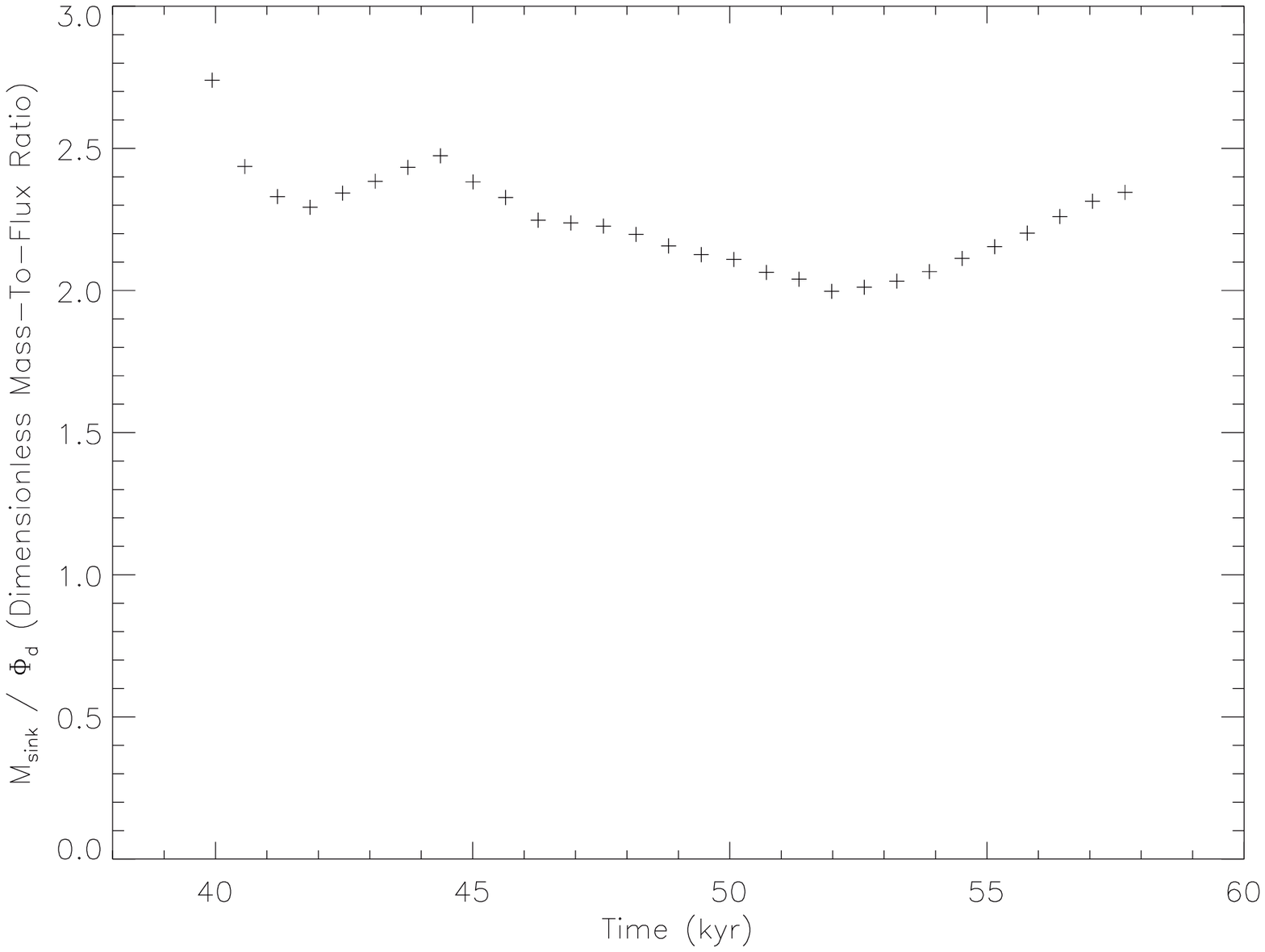}
\caption{The dimensionless ratio $\lambda_{\star,d}$ of the stellar mass $M_\star$ to the magnetic
  flux in the decoupling-enabled magnetic structure (DEMS) $\Phi_d$ as a
  function of time. The closeness of $\lambda_{\star,d}$ to the value
  $\lambda_{core}=2$ for the core as a whole indicates that the flux
  in the DEMS is released by the stellar material.}
\label{mtf2}
\end{figure}

We have checked that the creation and evolution of the DEMS is
insensitive to the details of the sink particle treatment. For
example, we have allowed 7 and 8 (instead of 6) levels of refinement before
the sink particle creation, which slowed down the computation
greatly. The results are qualitatively similar, although the shape and orientation of the DEMS are somewhat different. We have also varied the critical density $\rho_{crit}$ in the
equation of state, and obtained a broadly similar result. These 
tests support the conclusion that the DEMS is robust. Its existence
also makes physical sense because, as the decoupled flux accumulates
near the protostar, the magnetic pressure builds up, which can only be
released through expansion. In hindsight, it is hard to imagine any other 
outcome.

\subsubsection{DEMS and Collapse Dynamics}

%
%
The DEMS modifies the dynamics of core collapse and star formation in 
several ways. 
The most obvious is the change to the density distribution and velocity field in the equatorial
region where most of the mass accretion occurs. The DEMS 
produces an evacuated region that expands against the dense, 
collapsing pseudodisk, as shown in the left panel of Fig.~2. There are
initially several small, irregular, low-density regions. Only one
develops into a full-blown DEMS through runaway expansion. It starts to
grow quickly after enough magnetic flux has accumulated in
the region that a high magnetic pressure is built up to 
overwhelm the ram pressure of the collapsing flow. The expansion 
occurs presumably along the path of least resistance, which 
happens to lie in the direction towards the lower-left corner in
Fig.~\ref{magbub}; this direction is probably related to the cubic
simulation box and Cartesian grid, which break the symmetry in the collapsing flow.
As mentioned earlier, by the time $t=60$~kyrs (or 
about $2~t_{ff}$), the structure grows to a size comparable to the 
initial core radius, and starts to merge into the background.  


To examine the expansion dynamics more quantitatively, we plot in
Fig.~5 the distributions of the magnetic pressure, the
thermal pressure $P_{th}=\rho c_s^2$, and the ram pressure associated
with the radial component of the velocity $P_{r,ram}= \rho v_r^2$, in
the equatorial plane, along the dotted line shown in the left panel of
Fig.~2. Note the
sharp increase (by a factor of $\sim 10^3$) in the magnetic pressure 
around $r\sim 10^{16}$~cm, which marks the boundary of the
DEMS. Inside the boundary, the magnetic pressure is more or less
uniform. It is much higher than
the thermal pressure, by a factor of $\sim 10^2$, corresponding to 
a low plasma $\beta\sim 0.01$. The high magnetic pressure is what drives the 
region to expand. The expansion is slowed down by the ram pressure of
the collapsing material outside the strongly magnetized region, which
is somewhat smaller than, but comparable to, the magnetic pressure. 
The spike just outside $r=10^{16}$~cm on 
the curve of ram pressure is due to the dense, expanding ring that is
prominent in Fig.~2. The ring is the collapsing pseudodisk material
that is swept up by the expanding, magnetically dominated region. It
has a peak density nearly two orders of magnitude higher than the
surrounding medium. The ring is not yet massive enough in our current
simulation to become self-gravitating, but we speculate that this might 
happen 
under other circumstances; in such cases, the ring may fragment into 
secondary objects. We will postpone an exploration of this possibility
to a future investigation.  

\begin{figure}
\epsscale{1.0}
\plotone{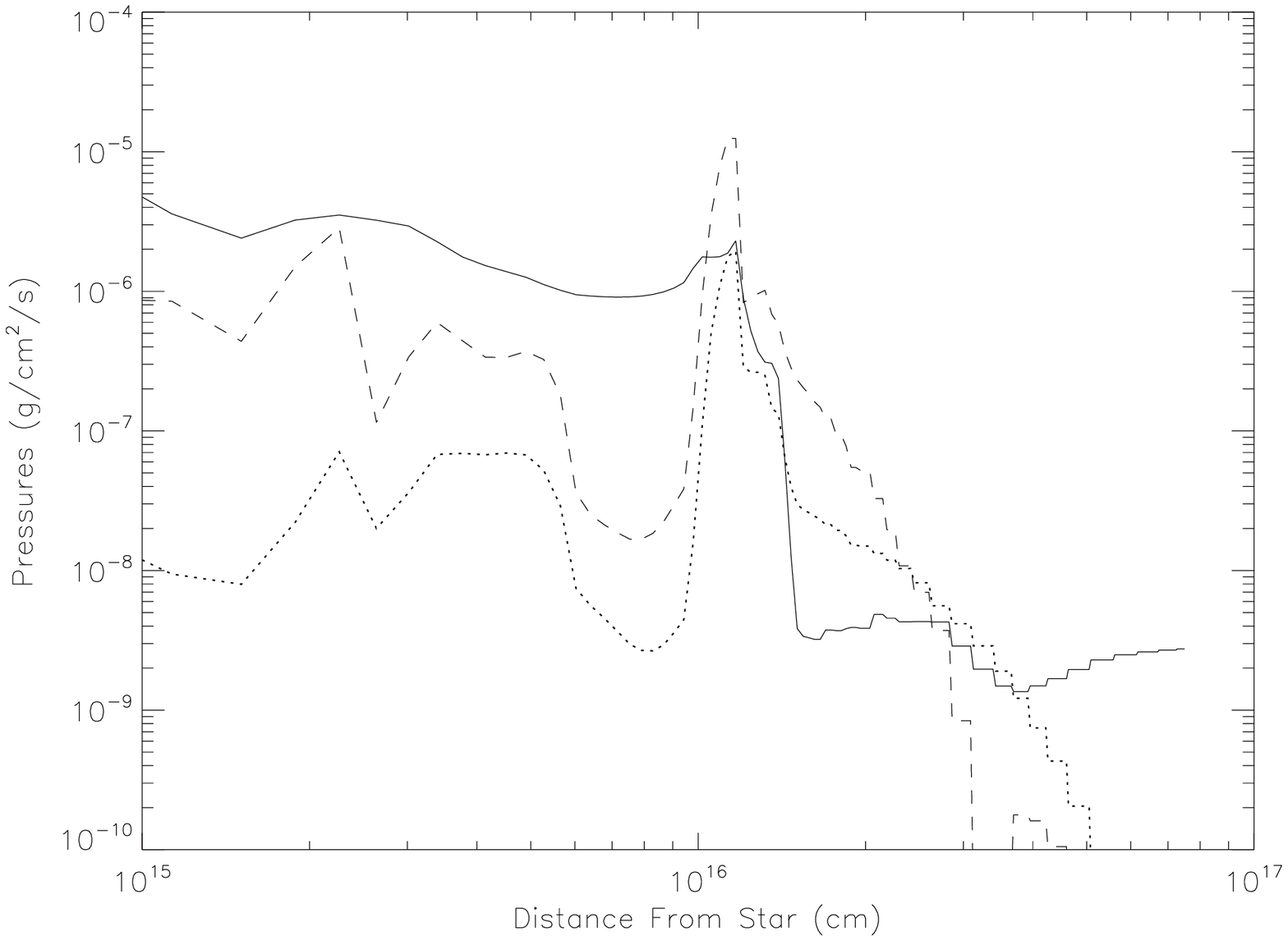}
\caption{Comparison of the magnetic (solid line), thermal (dotted) and
  ram pressure due to radial motion (dashed), in the equatorial plane,
  along a representative direction (shown as dotted line in the left
  panel of Fig.~\ref{magbub}). }
\label{profile2}
\end{figure}

A direct consequence of the one-sided expansion of the magnetically
dominated region is that accretion onto the central object must 
proceed in a highly asymmetric fashion. The asymmetric accretion is
shown vividly in the left panel of Fig.~2, where velocity vectors are
plotted in the equatorial plane of the star. While the collapsing flow
in the upper-right half of the plane can fall into the star directly,
that on the lower-left half is mostly obstructed by the DEMS. Based on
the highly asymmetric accretion pattern, one may expect the stellar
object to pick up some velocity. We find that the star particle in our
simulation does move, but only at a small speed of order 0.1~km/s, too 
small to be of any dynamical significance. The reason for the slow
stellar motion is probably the following: in the absence of any
external force, the total linear momentum of the system, which is
initially zero, must be conserved. In particular, the collapsing flow
on the lower-left side of the star is diverted by the DEMS to flow 
towards the star along the dense ring. Its momentum may cancel out
that of the unobstructed collapsing flow from the other side to a
large extent, at least in this particular simulation. Whether the slow
stellar motion is true in general remains to be determined.   

The diversion of the magnetized collapsing flow around the DEMS
creates an
interesting feature: the twisting of magnetic field lines, which is
clearly visible in Fig.~3. The twisting is normally not expected in a
non-rotating collapse. However, when the direct path to the star is
blocked by the DEMS, a parcel of collapsing flow moves around the
structure, twisting the field lines tied to the parcel in the
process. Indeed, the magnetic tension associated with the twisted
field lines produces an bipolar outflow moving away from the equatorial 
plane (most visible from the velocity field in the x-z or y-z plane,
not shown), similar to the cases where an initial core rotation is
present (e.g., \ct{Tomisaka1998}). Such an outflow has not been seen
in previous non-rotating collapse calculations. 

\subsection{Rotating Collapse}
\label{rotating}

We have shown that the decoupling-enabled magnetic structure strongly
affects the dynamics of non-rotating collapse. Here, we wish to
examine its influence on the dynamics of rotating collapse in general
and disk formation in particular. For this purpose, we impose on the
core an initial solid-body rotation with $\Omega=4.0 \times
10^{-13}$~s$^{-1}$, keeping other parameters the same as in the
non-rotating collapse discussed in \S~\ref{non-rotating}. 

Fig.~6 (left panel) shows a snapshot of the rotating collapse at a 
representative time $t=47.5$~kyrs (or $\sim 1.6~t_{ff}$), when the
stellar object has accreted $\sim 0.67\msun$. The overall morphology
is broadly similar to that of the non-rotating case shown in Fig.~2,
with an evacuated DEMS expanding against a collapsing flow. An obvious
difference is, of course, rotation. The rotation is not fast enough,
however, to stop the collapse centrifugally, even along directions not
directly affected by the DEMS. The lack of complete centrifugal
support is illustrated in the right panel of Fig.~6, where we plot the
infall and rotation speeds along a DEMS-free direction (marked in the
right panel as a dotted line). The infall speed remains well below the
Keplerian value, except in the region inside a radius of $\sim 8\times
10^{14}$~cm, where the two become comparable. In this small region,
the collapse is slowed down, but not stopped. Its infall speed is
$\sim 1.5-2$~km/s, much higher than the sound speed (0.2~km/s). The
supersonic infall leaves little doubt that a large-scale rotationally
supported structure is not formed in this direction, presumably
because of strong magnetic braking, which has been shown to be capable
of suppressing disk formation in previous ideal MHD simulations (\ct{Allen+2003}; \ct{MellonLi2008}; \ct{HennebelleFromang2008}).

\begin{figure}
\epsscale{1.2}
\plottwo{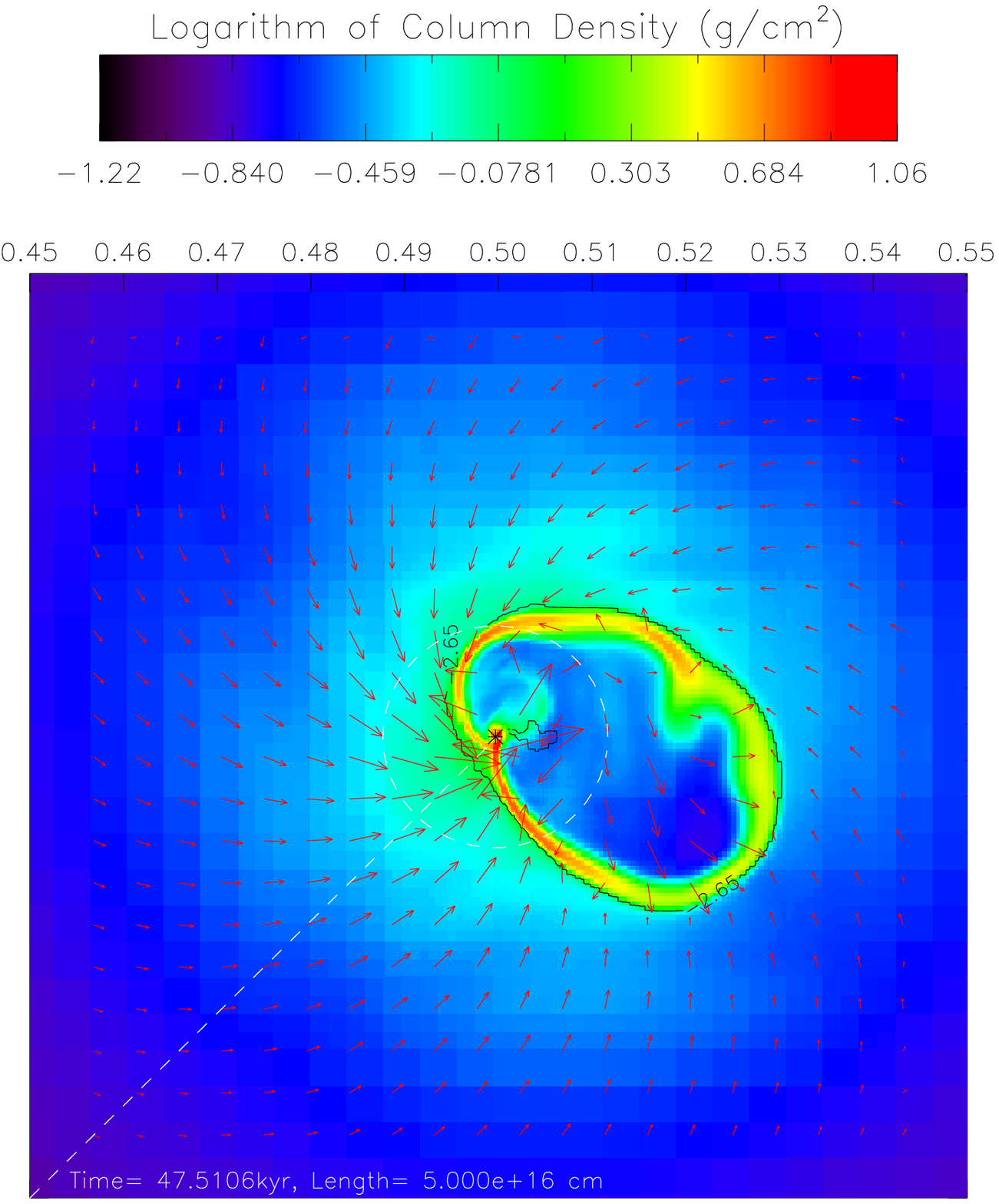}{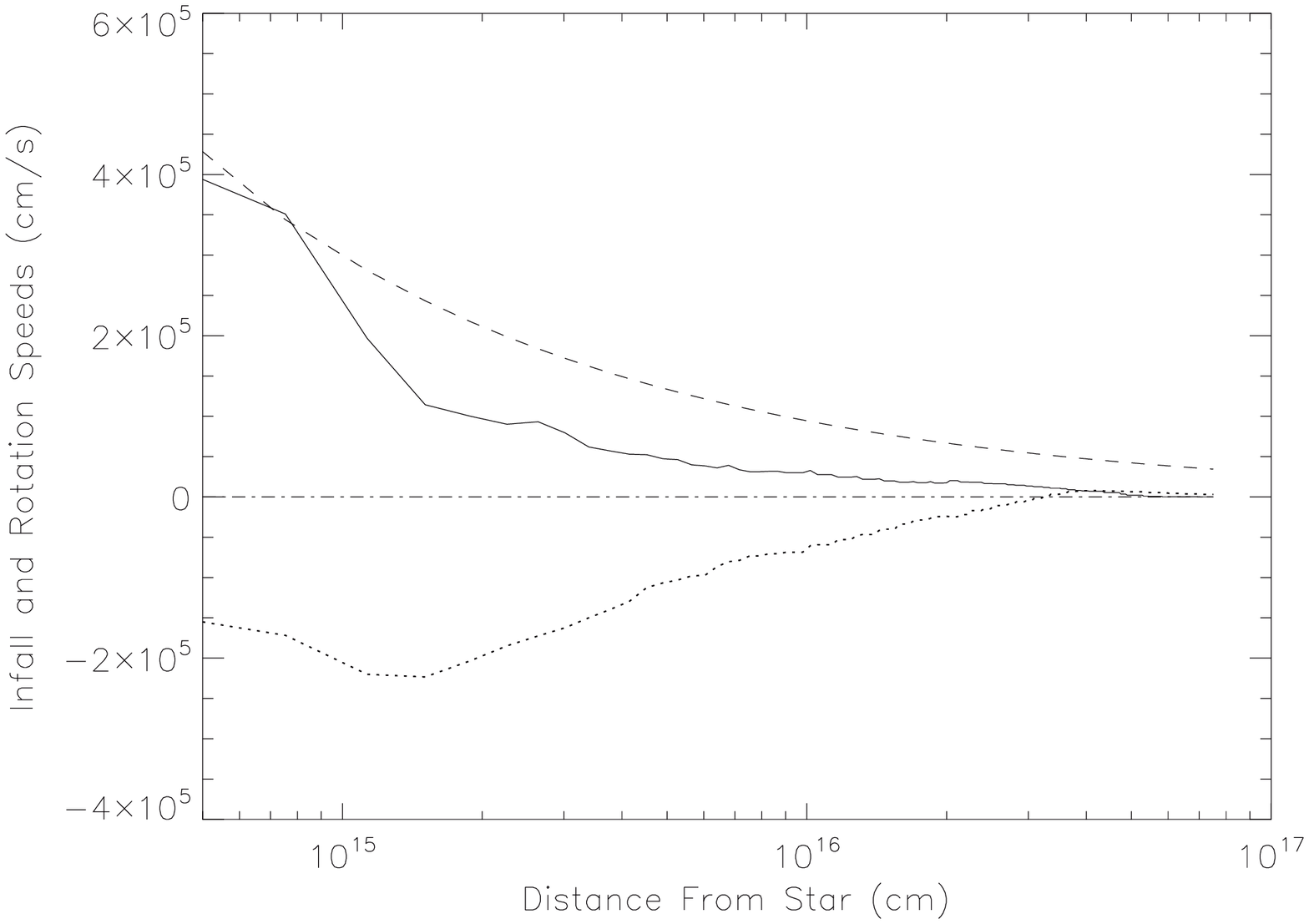}
\caption{Left panel: Same as the left panel of Fig.~\ref{magbub} but
  for rotating collapse. Again, an expanding low-density region is
  clearly visible. Right panel: Infall (dotted line) and rotation
  (solid) speed on the equatorial plane along the dotted line shown in
  the left panel. A Keplerian profile is also plotted (dashed) for
  comparison. 
}
\label{rot}
\end{figure}

The DEMS makes disk formation even more difficult. The reason is that
the DEMS is a rather rigid structure that prevents the rotating
material from completing a full orbit around the central star. To be
more quantitative, we compare the magnetic pressure 
to the ram pressure due to rotation 
$P_{\phi,ram}=\rho v_\phi^2$ along a circle of 400~AU in radius
in the equatorial plane in Fig.~7. Clearly, the magnetic pressure
inside the DEMS ($\sim 0^\circ - 100^\circ$ and $\sim 290^\circ -
360^\circ$) is higher than the rotational ram pressure outside the
DEMS (the two peaks of ram pressure correspond to locations on the dense ring). The DEMS is effectively a magnetic wall that stops the rotating motion of the material that runs into it. This is a new form of magnetic braking that has never been discussed before in core collapse and disk
formation. 

\begin{figure}
\epsscale{1.0}
\plotone{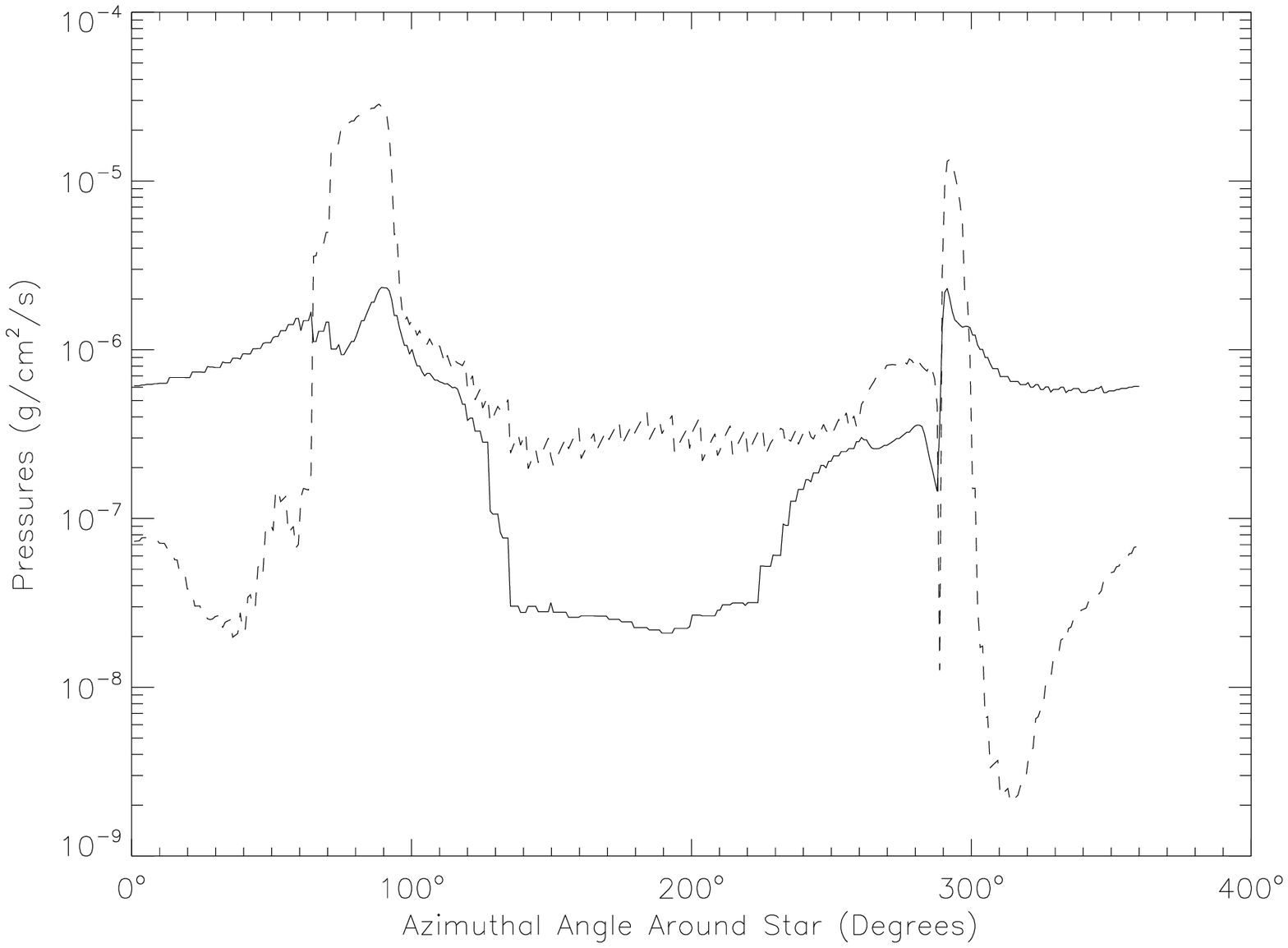}
\caption{Comparison of the magnetic pressure (solid line) and the ram
  pressure due to rotation in the equatorial plane, on a circle of 
$\sim 400$~AU in radius (shown in the left panel of
  Fig.~\ref{rot}). The azimuthal angle is measured counterclockwise
  from the x-axis. The magnetic pressure inside the DEMS ($\sim
  0^\circ-100^\circ$ and $\sim 290^\circ-360^\circ$) is much higher
  than the rotational ram pressure outside ($\sim 100^\circ -
  290^\circ$), making disk formation difficult. }
\label{rotpres}
\end{figure}

The rotation has an effect on the shape of the dense ring. Whereas the
ring is more or less (mirror) symmetric with respect to a plane in the
non-rotating collapse (see the dotted line in Fig.~2), it is much less
so in the rotating collapse. The reason is that one side of the DEMS
expands against the rotating flow, and its expansion is slowed down by 
rotation (see the lower side of the DEMS in the left panel of 
Fig.~6). The other side expands into a medium that is already 
rotating away from it to begin with, and its expansion is sped up by
rotation. The net result is that the lateral expansion of the DEMS
becomes asymmetric in the presence of rotation. The effect is particularly 
strong at early times, when the rotation speed has yet to be greatly
reduced by magnetic braking and the DEMS. 

\section{Discussion and Summary}
\label{discussion}


We find a new feature in the protostellar collapse of magnetized dense
cores: an expanding low-density region driven by the magnetic flux
decoupled from the material that has accreted onto the star. This decoupling-enabled magnetic structure (DEMS) has implications for both high resolution star formation simulations and for collapse dynamics including disk (and possibly binary) formation.

The DEMS formation is a natural consequence of simulating magnetized
star formation using the sink particle formalism. Sink particles are
needed to represent the formed stellar objects because such objects
are much smaller and much denser than their parental dense cores
(e.g., \ct{Krumholz+2004}). When the matter from a cell is added to a
sink particle, the magnetic flux from the cell cannot be added to the
sink as well, on both physical and numerical grounds. Physically, the
addition of the magnetic flux to the sink particle would make the
stellar field strength much higher than observed (which is, of course,
the well  known ``magnetic flux problem''). Numerically, the sink
particle cannot hold a large magnetic flux, which would produce a
large, unbalanced magnetic force in the host cell of the sink
particle. The needed decoupling of the magnetic field from matter 
during sink particle mass accretion makes the creation of the 
DEMS unavoidable in an ideal MHD simulation of the protostellar 
phase of star formation.

%
%
The DEMS that we found is conceptually related to the highly
magnetized inner region of protostellar collapse in the presence 
of non-ideal MHD effects, found 
previously either semi-analytically (in 1D) or through 2D 
(axisymmetric) simulations. It has been shown that, in the presence of 
ambipolar diffusion, most of the magnetic flux left 
behind by the stellar mass is trapped in a strongly magnetized 
region inside a C-shock (\ct{LiMcKee1996},
\ct{CiolekKonigl1998}). 
The situation is similar in the presence of Ohmic 
dissipation (\ct{Li+2011}). In both cases, because of the 
axisymmetry assumed, the collapsing flow has to cross the strongly
magnetized inner region where the left-over magnetic flux is parked 
in order to reach the center. As a result, the highly magnetized 
region is loaded with high-density material (at least near the 
equatorial plane, see, e.g., Fig.~4 of \ct{Li+2011}). 
Strictly speaking, the region is not a (low-density) DEMS, 
although it does share the same origin as the DEMS: both are driven by
the decoupled magnetic flux. 
Indeed, one may view the case considered in this paper as the
high-ionization limit of the general non-ideal case in which the ideal
MHD approximation breaks down only at the highest densities (and the
breakdown is mimicked here by the sub-grid physics associated with
sink particle treatment). In this limiting case, a very small region
of intense magnetic field is expected to form close to the protostar
and be trapped by the ram pressure of the infalling material in 2D. 
In 3D, we find a completely different behavior: an
asymmetrically expanding DEMS. The reason is that, in 3D, the
collapsing core material does not have to go {\it through} the strongly
magnetized region to reach the central object; it can simply go around
the region. The DEMS in 3D is able to choose a path of least
resistance (which may be influenced by grid geometry), breaking the
restrictions imposed in strictly axisymmtric simulations. 
%
%
%


The DEMS is a new feature never reported before in 3D numerical 
simulations of star formation. For example, 
\citet{HennebelleFromang2008} followed the collapse of magnetized 
cores using an AMR MHD code, but did not mention any structure 
similar to our DEMS. The reason for the absence of DEMS in their 
simulations is probably that they did not use any sink particle in 
their simulations and thus did not address the issue of magnetic 
decoupling that is needed to resolve the magnetic flux problem. 
\citet{Machida+2010} did use sink particles in their nested grid 
MHD simulations. There was no mention of any DEMS-like structure 
in their paper, however. It was not clear what happened to the 
magnetic flux associated with the stellar mass, especially for 
their formally ideal MHD simulations where the magnetic flux is expected 
to be conserved.
%
%


%
%

Although we believe that the DEMS is robust in the limit considered 
in this paper where the matter and magnetic field are well coupled 
except at the highest densities, it remains unclear how it will 
be affected by non-ideal MHD effects, including ambipolar 
diffusion, Ohmic dissipation and Hall effect, all of which can 
play a role in the magnetic field evolution during core collapse 
and disk formation, at least in 2D. The logic next step is to 
carry out 3D AMR MHD simulations that include both sink 
particles and non-ideal MHD effects. It would be interesting 
to determine the extent to which the collapsing flow reaches 
the star either by crossing field lines in a strongly magnetized 
region through non-ideal MHD effects (as in the current 2D 
simulations, such as \citet{Li+2011}) or by going around the 
strongly magnetized region, as we find in this paper.

To summarize, we have carried out simulations of magnetized core
collapse and star formation using an MHD version of the ENZO 
AMR code. A sink particle treatment is used to decouple the magnetic
flux from the material that enters the star, which must happen to
resolve the well-known ``magnetic flux problem'' in star formation. 
We find that the decoupled flux creates a low-density high magnetic
pressure region that expands anisotropically away from the
protostar. This decoupling-enabled magnetic structure has profound
effects on the dynamics of core collapse, making the protostellar 
accretion flow highly asymmetric and the formation of a rotationally 
supported disk more difficult. It is a generic feature of star
formation in magnetized cloud cores that should be included in 
future simulations, especially those using sink particle treatment.

\acknowledgments
We thank Peng Wang for help with the ENZO MHD code used in this
work. It was supported in part by NASA through NNG06GJ33G and
NNX10AH30G, by the Theoretical Institute for Advanced Research in
Astrophysics (TIARA) under the CHARMS initiative and the National Science Council of Taiwan through grant NSC97-2112-M-001-018-MY3, and by Scientific Research of Japan through Grant-in-Aid 20540228 and 22340040.

\end{document}